\documentclass[12pt]{iopart}
\usepackage{graphicx}
\usepackage{setstack,cite}
\usepackage{iopams}

\begin{document}
%\eqnobysec

\title[Completely Integrable $N$-Coupled Li\'enard Type Nonlinear Oscillator]
{Dynamics of a Completely Integrable $N$-Coupled Li\'enard Type Nonlinear Oscillator}

\author{R Gladwin Pradeep, V K Chandrasekar, M Senthilvelan and M Lakshmanan}
\address{Centre for Nonlinear Dynamics, School of Physics,
Bharathidasan University, Tiruchirappalli - 620 024, India }
\ead{lakshman@cnld.bdu.ac.in}

%\date{\today}

\begin{abstract}
We present a system of $N$-coupled Li\'enard type nonlinear
oscillators which is completely integrable and possesses explicit $N$ time-independent
and $N$ time-dependent integrals.  In a special case, it becomes maximally
superintegrable and admits $(2N-1)$ time-independent integrals.  The results are illustrated
for the $N=2$ and arbitrary number cases.  General explicit
periodic (with frequency independent of amplitude) and quasiperiodic solutions as well as decaying type/frontlike
solutions are presented, depending on the signs and magnitudes of the system parameters.  Though
the system is of a nonlinear damped type, our investigations show that it possesses a Hamiltonian structure and that under
a contact transformation it is transformable to a system of uncoupled harmonic oscillators.
\end{abstract}

%\submitto{\JPA}

\maketitle
\section{Introduction}
In a recent paper we have shown that the modified Emden type equation (MEE) with additional
linear forcing,
\begin{eqnarray}
\ddot{x}+3kx\dot{x}+k^2x^3+\lambda x=0,\label{1dmee}
\end{eqnarray}
where over dot denotes differentiation with respect to $t$ and $k$ and $\lambda$ are arbitrary
parameters, exhibits certain unusual nonlinear dynamical properties \cite{pre}. Equation (\ref{1dmee})
is essentially of Li\'enard type.  For a particular
sign of the control parameter, namely $\lambda > 0$, the frequency of oscillations of the
nonlinear oscillator (\ref{1dmee}) is completely independent of the amplitude and remains the same as
that of the linear harmonic oscillator, thereby showing that the \emph{amplitude dependence of
frequency is not necessarily a fundamental property of nonlinear dynamical phenomena.}
In this case ($\lambda>0$) the system admits the explicit sinusoidal periodic
solution
\begin{eqnarray}
x(t)=\frac{A\sin(\omega t+\delta)}{1-\left(\frac{k}{\omega}\right)A
\cos(\omega t+\delta)},\quad 0\le A<\frac{\omega}{k},\quad \omega=\sqrt{\lambda},\label{sinusoidalsol}
\end{eqnarray}
where $A$ and $\delta$ are arbitrary constants.
\begin{figure}[!ht]
\begin{center}
\includegraphics[width=.8\linewidth]{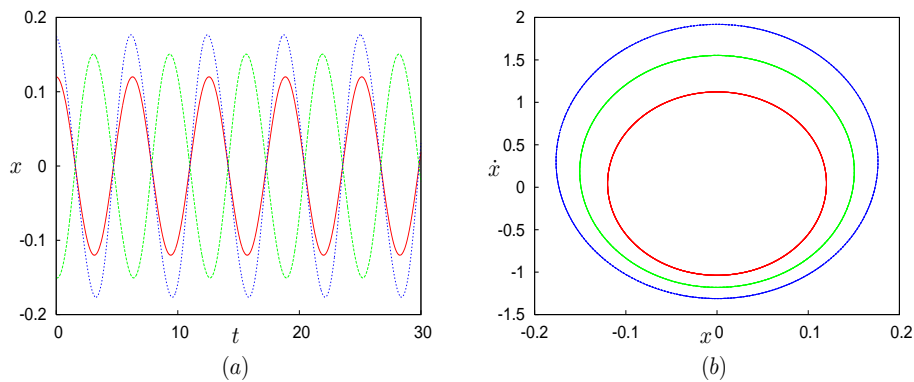}
\caption{(Color online) Solution and phase space plots of equation (\ref{1dmee}) for the case $\lambda<0$ (a) Periodic oscillations (b) Phase space portrait
}
\label{fig1}
\end{center}
\end{figure}
 In figure \ref{fig1}a we depict the harmonic
periodic oscillations of the MEE (\ref{1dmee}) for three different initial conditions, showing the amplitude
independence of the period or frequency.  The phase space plot, figure \ref{fig1}b, resembles
that of the
harmonic oscillator which again confirms that the system has a unique period of
oscillations for $\lambda>0$.

For $\lambda<0$, equation (\ref{1dmee}) admits the following form of solution \cite{pre},
\begin{eqnarray}
x(t)=\left(\frac{\sqrt{|\lambda|}(I_1\e^{2\sqrt{|\lambda|}t}-1)}
{kI_1I_2e^{\sqrt{|\lambda|}t}+k(1+I_1e^{2\sqrt{|\lambda|}t})}\right),\label{decayingsol}
\end{eqnarray}
where $I_1$ and $I_2$ are constants.  Depending on the initial condition, the
solution (\ref{decayingsol}) is either of decaying type
 or of aperiodic frontlike type, see figure 2a.
The time of decay or approach to asymptotic value is independent of the amplitude/initial value, which is once again
an unusual feature for a nonlinear dynamical system \cite{pre}.
\begin{figure}[!ht]
\begin{center}
\includegraphics[width=.8\linewidth]{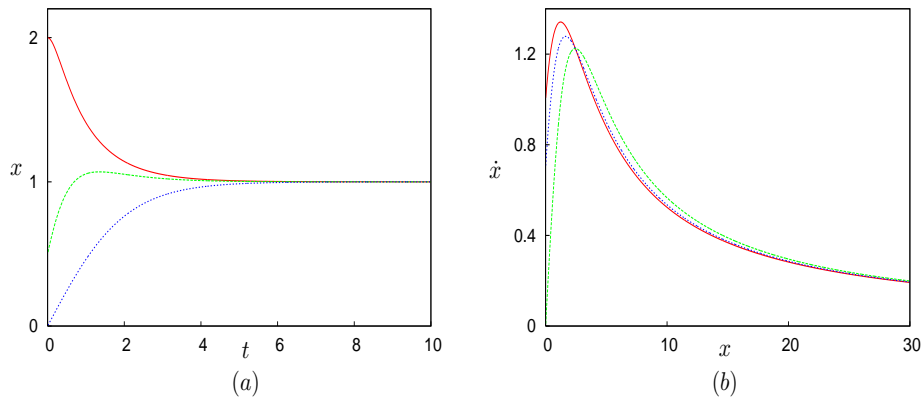}
\caption{(Color online) (a) Decaying and frontlike solutions of (\ref{1dmee}) for the parametric choice $\lambda<0$, (b) Solution plot of (\ref{1dmee}) with $\lambda=0$}
\end{center}
\end{figure}
Finally, for $\lambda=0$ equation (\ref{1dmee}) is nothing but the MEE \cite{emden} which has the exact general solution \cite{leach}
\begin{eqnarray}
x(t)=\frac{t+I_1}{2kt^2+I_1kt+I_2},
\end{eqnarray}
where $I_1$ and $I_2$ are the two integrals of motion (figure 2b).

A natural question which now arises is whether there exist  higher dimensional
coupled analogues of (\ref{1dmee}) which are integrable and exhibit
interesting oscillatory properties.  In this paper we first report a system of two-coupled
MEEs (with additional linear forcing) which is completely integrable, possesses two time-independent integrals
and two time-dependent integrals and whose general solution can be
obtained explicitly.  Depending on the signs of the linear term, the system admits
 periodic (with amplitude independent frequency) or quasiperiodic solutions or bounded aperiodic solutions
 (decaying or frontlike).  The results are then extended
to $N$-coupled MEEs and we
prove the complete integrability of them also in the same way.  From the nature of the explicit solutions we identify a
suitable contact transformation which maps the coupled system
 onto a system of coupled
canonical equations corresponding to a system of $N$-uncoupled harmonic oscillators, thereby proving
the Hamiltonian nature of the $N$-coupled MEEs.
We also prove that
the system becomes a maximally superintegrable one \cite{tempesta} for any value of $N(>1)$ when the coefficients of
the linear force
terms are equal.

We organize our results as follows.  In Section 2 we propose a two-coupled version of the MEE (\ref{1dmee}) and
construct the integrals of motion through the recently proposed modified Prelle-Singer procedure (a brief description of this
procedure is given in Appendix A).
In Section 3, by appropriately choosing  the coefficients of the linear forcing term,  we construct different forms of
general solutions including periodic, quasiperiodic and frontlike solutions and discuss the dynamics in each of the
cases in some detail.  In Section 4, we briefly analyse the symmetry and Painlev\'e singularity structure properties of the proposed
two-coupled version of the MEE.  We also investigate the dynamics of this equation under perturbation by numerical analysis.
We consider a $N$- coupled generalization of equation (\ref{1dmee}) and discuss
the general solution/dynamics in Section 5.  In Section 6, we identify a Hamiltonian structure for the $N$-coupled MEE
by mapping it onto a system of $N$-uncoupled
 harmonic oscillators through a contact type transformation which is obtained from the general solution of the coupled MEE.
Finally in Section 7, we summarize our results.  In Appendix A, we briefly discuss the generalized modified Prelle-Singer procedure
which has been used to derive the results discussed in Section 2.
\section{Two-dimensional generalizations of the MEE}
It is of considerable interest to study the dynamics of higher dimensional versions
of the MEE (\ref{1dmee}).  Recently Cari\~nena and Ranada  studied the uncoupled two-dimensional
version of equation (\ref{1dmee}) \cite{carinena},
\begin{eqnarray}
&&\ddot{x}+3k_1x\dot{x}+k_1^2x^3+\lambda_1x=0,\nonumber\\
&&\ddot{y}+3k_2y\dot{y}+k_2^2y^3+\lambda_2 y=0,\label{uncoupledmee}
\end{eqnarray}
where $k_1,\,k_2,\,\lambda_1$ and $\lambda_2$ are arbitrary parameters,
and analyzed the geometrical properties and proved that the above system
is superintegrable \cite{carinena}.
On the other hand, Ali et al.  \cite{mahomed}  have
 analysed a system of two-coupled differential equations, which is a complex version
of (\ref{1dmee}) with $x=y+iz$ and $\lambda=0$:
\begin{eqnarray}
&&\ddot{y}=-3(y\dot{y}-z\dot{z})-(y^3-3yz^2),\nonumber\\
&&\ddot{z}=-3(z\dot{y}+y\dot{z})-(3y^2z-z^3).\label{complexcmee}
\end{eqnarray}
The above system of equations is shown to be linearizable by complex point transformation
and from the solution of the linearized equation, a general solution for the equation (\ref{complexcmee}) has been constructed \cite{mahomed}.

In \cite{kundu} we have pointed out that equation (\ref{1dmee}), and its generalization as well as the $N^{th}$
order version can be
transformed to linear differential equations through appropriate nonlocal transformations.
In particular
equation (\ref{1dmee}) under the nonlocal transformation $U=x(t)e^{\int_0^tkx(\tau)d\tau}$ gets transformed
to the linear harmonic oscillator equation
 $\ddot{U}+\lambda U=0$, where $x$ and $U$ are also related through the Riccati equation $\dot{x}=\frac{\dot{U}}{U}x-kx^2$.
Substituting the expressions for $U$ and $\dot{U}$ and solving the resultant Riccati equation one can obtain the
solution (\ref{sinusoidalsol}).  Now searching for
possible extensions to higher dimensions by considering a generalized nonlocal transformation of the
form $U=xe^{\int_0^tf(x(\tau),y(\tau))d\tau}$,
$V=ye^{\int_0^tg(x(\tau),y(\tau))d\tau}$, where $U$ and $V$ satisfy the uncoupled linear harmonic oscillator equations,
$\ddot{U}+\lambda_1 U=0$ and $\ddot{V}+\lambda_2 V=0$, we try to identify the forms $f(x,y)$ and $g(x,y)$ so that the
transformation can be written as a system
of parametrically driven Lotka-Volterra type equations or time-dependent coupled Riccati equations.  In particular, with
the choice $f=g=k_1x+k_2y$, the transformation becomes a set of coupled time dependent Riccati equations,
$\dot{x}=\left(\frac{\dot{U}}{U}x-k_1x^2-k_2xy\right)$, $\dot{y}=\left(\frac{\dot{V}}{V}y-k_1xy-k_2y^2\right)$.  Consequently one obtains a system of two coupled MEEs with additional
linear forcing,
\begin{eqnarray}
&&\ddot{x}=-2(k_1x+k_2y)\dot{x}-(k_1\dot{x}+k_2\dot{y})x-
(k_1x+k_2y)^2x-\lambda_1 x\equiv\phi_1,\nonumber\\
&&\ddot{y}=-2(k_1x+k_2y)\dot{y}-(k_1\dot{x}+k_2\dot{y})y-
(k_1x+k_2y)^2y-\lambda_2 y\equiv\phi_2,\label{secondordercmee}
\end{eqnarray}
where $k_i$'s and $\lambda_i$'s, $i=1,2,$ are arbitrary parameters.
When either one of the parameters $k_1$ or $k_2$ is taken as zero, then one of the two equations in (\ref{secondordercmee})
reduces to the MEE defined by (\ref{1dmee})  while the other reduces to a linear ordinary differential equation (ODE) in the other variable or vice versa.  On the other hand
when one of the variables ($x$ or $y$) is zero, equation (\ref{secondordercmee}) reduces to a MEE in the other
variable.  A characteristic feature
of this form (\ref{secondordercmee}) is that it can be straightforwardly extended to higher dimensions as
we see in the following sections besides admitting unusual nonlinear dynamical properties.

%\section{Second-order coupled modified Emden type equation}
To obtain the solutions of the above system of nonlinear ODEs one can solve the above coupled Riccati equations. However,
to obtain the integrals of motion as well as the solutions we find it more convenient to solve (\ref{secondordercmee}) by the generalized modified
Prelle-Singer (PS) procedure introduced recently \cite{royal1}.  We indicate this procedure applicable to
(\ref{secondordercmee}) briefly in Appendix A.   The resultant independent integrals of motion can be
written as
\begin{eqnarray}
&&\hspace{-1.2cm}I_1=\frac{(\dot{x}+(k_1x+k_2 y)x)^2+\lambda_1 x^2}
{\left[\frac{k_1}{\lambda_1}(\dot{x}+(k_1x+k_2y)x)+\frac{k_2}{\lambda_2}(\dot{y}+(k_1x+k_2y)y)+1\right]^2},\label{2dtidint1}\\
&&\hspace{-1.2cm}I_2=\frac{(\dot{y}+(k_1x+k_2 y)y)^2+\lambda_2 y^2}
{\left[\frac{k_1}{\lambda_1}(\dot{x}+(k_1x+k_2y)x)+\frac{k_2}{\lambda_2}(\dot{y}+(k_1x+k_2y)y)+1\right]^2},\label{2dtidint2}\\
%\end{eqnarray}
%\begin{eqnarray}
&&\hspace{-1.2cm}I_3=\left\{
\begin{array}{ll}
\tan^{-1}\left[\frac{\sqrt{\lambda_1}x}{\dot{x}+(k_1x+k_2y)x}\right]
-\sqrt{\lambda_1}\,t,&\lambda_1>0\label{2dtdint1}\\
\frac{e^{2\sqrt{|\lambda_1|}t}(\dot{x}+(k_1x+k_2y)x-\sqrt{|\lambda_1|}x)}
{\dot{x}+(k_1x+k_2y)x+\sqrt{|\lambda_1|}x},&\lambda_1<0,\label{2dtdint3}
\end{array}
\right.\\
&&\hspace{-1.2cm}I_4=\left\{
\begin{array}{ll}
\tan^{-1}\left[\frac{\sqrt{\lambda_2}y}{\dot{y}+(k_1x+k_2y)y}\right]
-\sqrt{\lambda_2}\,t,&\lambda_2>0\label{2dtdint2}\\
\frac{e^{2\sqrt{|\lambda_2|}t}(\dot{y}+(k_1x+k_2y)x-\sqrt{|\lambda_2|}y)}
{\dot{y}+(k_1x+k_2y)y+\sqrt{|\lambda_2|}y},&\lambda_2<0.\label{2dtdint4}
\end{array}
\right.
\end{eqnarray}

We note here that the forms of the time-dependent integrals of motion depend upon the signs of the parameters
$\lambda_i$, $i=1,2$.  Using the above integrals one can obtain periodic and aperiodic but bounded solutions as per (i) $\lambda_1,\,\lambda_2>0$
 (ii) $\lambda_1,\lambda_2< 0$ and (iii) $\lambda_1<0,\,\lambda_2>0$ (or vice versa).  The case $\lambda_1=\lambda_2=0$ is dealt with separately below in Section 3.3.  In the following we discuss the nature of the solutions.

\section{The Dynamics}
\subsection{Periodic and quasi periodic oscillations $(\lambda_1,\lambda_2>0)$}
 By restricting $\lambda_1,\,\lambda_2>0$
in the integrals (\ref{2dtidint1})-(\ref{2dtdint2}) and solving them algebraically we obtain the
following general solution,
\begin{eqnarray}
&&\hspace{-0.8cm}x(t)=\frac{A\sin(\omega_1 t+\delta_1)}
{1-\frac{Ak_1}{\omega_1}\cos(\omega_1 t+\delta_1)
-\frac{Bk_2}{\omega_2}\cos(\omega_2 t+\delta_2)},\nonumber\\
&&\hspace{-0.8cm}y(t)=\frac{B\sin(\omega_2 t+\delta_2)}
{1-\frac{Ak_1}{\omega_1}\cos(\omega_1 t+\delta_1)
-\frac{Bk_2}{\omega_2}\cos(\omega_2 t+\delta_2)},\,\,\quad
\bigg|\frac{Ak_1}{\omega_1}+\frac{Bk_2}{\omega_2}\bigg|<1,\label {cmee04b}
\end{eqnarray}
where $\omega_j=\sqrt{\lambda_j},\;j=1,2,\,A=\sqrt{I_1}/\omega_1,\,\,
B=\sqrt{I_2}/\omega_2,\,\delta_1=I_3,\,\,\delta_2=I_4$.  Two types of oscillatory motion
can arise depending on whether the ratio $\omega_1/\omega_2$ is rational or irrational
leading to $m:n$ periodic or quasiperiodic motion, respectively.  One may note that the frequency of
oscillations is again independent of the amplitude in the present two-coupled generalization also.
The conservative nature of the above oscillatory solution can
be seen from the phase space plot.
\begin{figure}[!ht]
\begin{center}
\includegraphics[width=.8\linewidth]{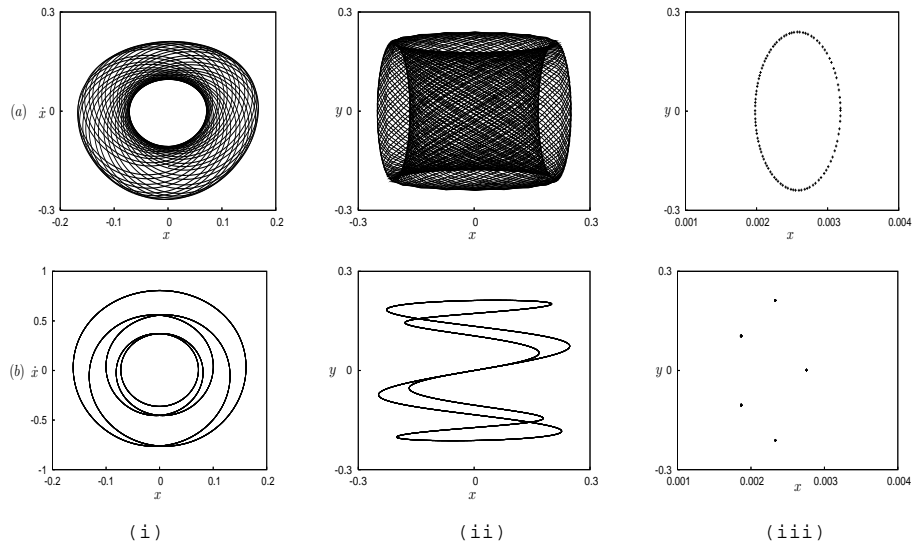}
\caption{(a) Quasi-periodic oscillations with $\omega_1=1$ and $\omega_2=\sqrt{2}$
(i) phase space plot (ii) Configuration space plot and (iii) Poincar\'e SOS
(b) periodic oscillations with $\omega_1=1$ and $\omega_2=5$
(i) phase space plot (ii) Configuration space plot and (iii) Poincar\'e SOS, see equation (\ref{cmee04b}). }
\end{center}
\label{fig2}
\end{figure}

To visualize the dynamics we plot the solutions $x(t)$ and $y(t)$ both in phase space and in
configuration space for two different sets of frequencies.  First we consider the case in which the ratio of
frequencies is an irrational number.  For illustration, we take $\omega_1=\omega_2/\sqrt{2}=1$.  The $(x,\dot{x})$
phase space plot is depicted in figure  3a(i).  The
configuration space $(x,y)$ plot is given in figure 3a(ii).  We also confirm the quasi-periodic nature of the
oscillations by
plotting the Poincar\'e surface of section (SOS) \cite{mlbook} in figure 3a(iii).  The solution is clearly not a closed
orbit and is
quasiperiodic or almost periodic in the sense that trajectory returns arbitrarily close to its starting point
infinitely often, which is confirmed by a closed curve in the Poincar\'e SOS. The system parameters are chosen
as $k_1=-1,k_2=1$ and the values of the arbitrary constants are taken as $A=-1$ and $B=-0.4$.

Next, when the ratio of frequencies is a rational number one has periodic solutions.  In figure 3b we
show a 1:5 periodic solution by choosing $\omega_1=1$ and $\omega_2=5$.

\subsection{Aperiodic solutions $(\lambda_1,\lambda_2<0)$}
We next consider the case $\lambda_1<0,\,\lambda_2<0$.  In this case one obtains aperiodic
but bounded frontlike or decaying type solutions.  To see this, we use the two time-independent integrals
($I_1,I_2$) and time-dependent integrals ($I_3,I_4$) vide equations (\ref{2dtidint1}),(\ref{2dtidint2}),(\ref{2dtdint3})
 and (\ref{2dtdint4}), and obtain the following general
solution for equation (\ref{secondordercmee}) when $\lambda_1,\lambda_2<0$,
\begin{eqnarray}
&&\hspace{-2cm}x(t)=\frac{\lambda_2\sqrt{|\lambda_1|}(\mathfrak{I}_1e^{\sqrt{|\lambda_1|}t}
-\mathfrak{I}_3e^{-\sqrt{|\lambda_1|}t})}{k_1\lambda_2(\mathfrak{I}_1e^{\sqrt{|\lambda_1|}t}
+\mathfrak{I}_3e^{-\sqrt{|\lambda_1|}t})+k_2 \lambda_1(\mathfrak{I}_2e^{\sqrt{|\lambda_2|}t}
+\mathfrak{I}_4e^{-\sqrt{|\lambda_2|}t})-2},\nonumber\\
&&\hspace{-2cm}y(t)=\frac{\lambda_1\sqrt{|\lambda_2|}(\mathfrak{I}_2e^{\sqrt{|\lambda_2|}t}
-\mathfrak{I}_4e^{-\sqrt{|\lambda_2|}t})}
{k_1 \lambda_2
(\mathfrak{I}_1e^{\sqrt{|\lambda_1|}t}+\mathfrak{I}_3e^{-\sqrt{|\lambda_1|}t})
+k_2\lambda_1(\mathfrak{I}_2e^{\sqrt{|\lambda_2|}t}
+\mathfrak{I}_4e^{-\sqrt{|\lambda_2|}t})-2},\label{sol2}
\end{eqnarray}
where $\mathfrak{I}_1=\sqrt{I_1}/(\sqrt{I_3}\lambda_1\lambda_2)$,\,\,
$\mathfrak{I}_2=\sqrt{I_2}/(\sqrt{I_4}\lambda_1\lambda_2)$,\,\,
$\mathfrak{I}_3=\sqrt{I_1I_3}/(\lambda_1\lambda_2)$,\,\,
$\mathfrak{I}_4=\sqrt{I_2I_4}/(\lambda_1\lambda_2)$.
Here also we observe that the solution is of the same form as in the one dimensional case.  The decaying
type
solution is plotted in figure 4a and the frontlike solution is shown in figure 4b.
\begin{figure}[!ht]
\begin{center}
\includegraphics[width=1\linewidth]{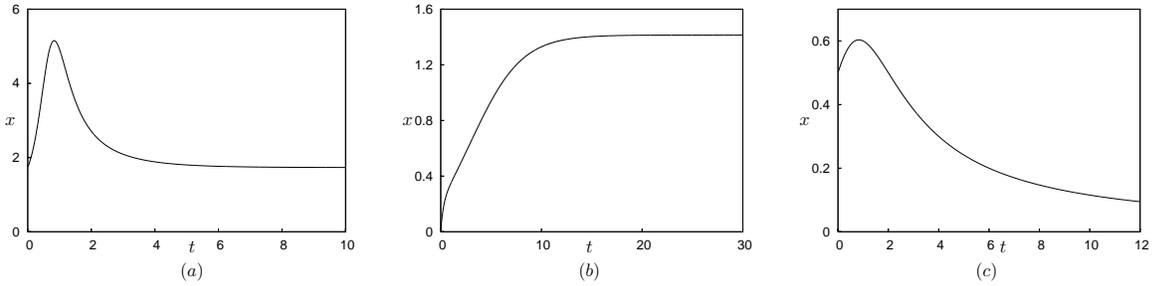}
\caption{(a) Decaying type solution of (\ref{secondordercmee}) for $\lambda_1,\,\lambda_2<0$ (b) Frontlike solution of (\ref{secondordercmee})
for $\lambda_1,\lambda_2<0$ (c) Decaying type solution
of (\ref{secondordercmee}) for $\lambda_1=\lambda_2=0$.}
\end{center}
\end{figure}
\subsection{Decaying type solution for $\lambda_1=\lambda_2=0$}
Restricting the values of the parameters to $\lambda_1=\lambda_2=0$
in (\ref{secondordercmee}) one
can get the two-dimensional generalization of the MEE (\ref{1dmee}) .
In this case we find that the system (\ref{secondordercmee}) admits the following
integrals of motion, namely,
\begin{eqnarray}
&\hspace{-1cm}I_1=\frac{(\dot{x}+(k_1x+k_2y)x)}
{\dot{y}+(k_1x+k_2y)y},\quad\,\,\, &I_2=-t+\frac{x}{\dot{x}+(k_1x+k_2y)x},\nonumber\\
&\hspace{-1cm}I_3=-t+\frac{y}{\dot{y}+(k_1x+k_2y)y},\,\,&I_4=\frac{t^2}{2}+
\frac{1-t(k_1x+k_2y)}{(k_1\dot{x}+k_2\dot{y}+(k_1x+k_2y)^2)^2}.
\end{eqnarray}

Then the general solution for equation (\ref{secondordercmee})
(with $\lambda_1=\lambda_2=0$) can be written in the form
\begin{eqnarray}
&&x(t)=\frac{2I_1(I_2+t)}{k_1I_1(2I_4+(2I_2+t)t)+k_2(2I_4+(2I_3+t)t)},\nonumber\\
&&y(t)=\frac{2(I_3+t)}{k_1I_1(2I_4+(2I_2+t)t)+k_2(2I_4+(2I_3+t)t)}.
\end{eqnarray}
Again the solution turns out to be a rational function in $t$ in which the
denominator is a quadratic function of $t$ as in the one dimensional case.
Choosing $I_1,\,I_2,\,I_3$ and $I_4$ suitably, one can obtain
 decaying type solutions as shown in figure 4c.

\subsection{Cases with mixed signs of parameters $\lambda_1$ and $\lambda_2$}
In the mixed case, for example with
 $\lambda_1>0,\,\lambda_2<0$, the solution can be derived from (\ref{sol2}) as
\begin{eqnarray}
&&\hspace{-1cm}x(t)=\frac{-\omega_1A\sin(\omega_1t+\delta_1)}{\frac{k_2\omega_1^2}{2}(\mathfrak{I}_2
e^{\sqrt{|\lambda_2|}t}
+\mathfrak{I}_4e^{-\sqrt{|\lambda_2|}t})+Ak_1\cos(\omega_1t+\delta_1)-1},\,\,\,\omega_1=\sqrt{\lambda_1},\nonumber\\
&&\hspace{-1cm}y(t)=\frac{\omega_1^2\sqrt{|\lambda_2|}(\mathfrak{I}_2e^{\sqrt{|\lambda_2|}t}-\mathfrak{I}_4
e^{-\sqrt{|\lambda_2|}t})}
{2[\frac{k_2\omega_1^2}{2}(\mathfrak{I}_2e^{\sqrt{|\lambda_2|}t}
+\mathfrak{I}_4e^{-\sqrt{|\lambda_2|}t})+Ak_1\cos(\omega_1t+\delta_1)-1]},\label{mixedsol}
\end{eqnarray}
where $A=\sqrt{I_1}/\omega_1^2$.

The motion now turns out to be a mixed oscillatory-bounded frontlike one.
We depict the solution (\ref{mixedsol}) in figure 5 in which we have fixed $k_1=k_2=1$, $\omega_1=2$, and $\lambda_2=-1$.
\begin{figure}[!ht]
\begin{center}
\includegraphics[width=.8\linewidth]{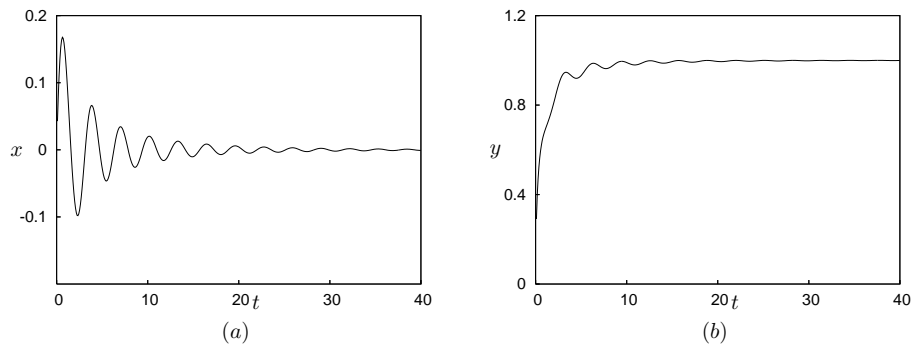}
\caption{Solution plots of equation (\ref{secondordercmee}) for the mixed case $\lambda_1>0$,  $\lambda_2<0$ (a) $x(t)$, (b) $y(t)$}
\end{center}
\end{figure}
\subsection{A superintegrable case}
%\begin{eqnarray}
%&&R=-\frac{k_1(2x(\dot{y}+k_2y^2)-\dot{x}y)+k_1^2x^2y+y(k_2(\dot{y}+k_2y^2)+\lambda)}
%{\lambda^2(k_1^2x^2+k_1(\dot{x}+2k_2xy)+k_2(\dot{y}+k_2y^2)+\lambda)^3}\\
%&&K=\frac{k_1^2x^3+k_1x(\dot{x}+2k_2xy)+k_2(2\dot{x}y+k_2xy^2-x\dot{y})+\lambda x}
%{\lambda^2(k_1^2x^2+k_1(\dot{x}+2k_2xy)+k_2(\dot{y}+k_2y^2)+\lambda)^3}
%\end{eqnarray}
Finally we investigate the dynamics in the limit $\lambda_1=\lambda_2=\lambda\ne0$.  In
this case one is able to find an additional time-independent integral of motion given by
\begin{eqnarray}
I_5=\frac{x\dot{y}-y\dot{x}}
{(k_1\dot{x}+k_2\dot{y}+(k_1x+k_2y)^2+\lambda)^2}.
\end{eqnarray}
As a result one has three time-independent integrals, $I_1,\,I_2,$ and $I_5$,
besides one time-dependent integral (any one of the two time-dependent integrals $I_3$ or $I_4$), in a two degrees of freedom system
which in turn confirms that
the system under consideration is a superintegrable one.  Obviously it is also maximally superintegrable \cite{tempesta}.
\section{Symmetry and Singularity structure analysis}
In order to understand why the system (\ref{secondordercmee}) is integrable and whether
perturbations of them lead to chaos, we analyse the symmetry properties, in particular the point
symmetries and Painlev\'e singularity structure associated with equation (\ref{secondordercmee}) and investigate
numerically the perturbed equation.
\subsection{Lie point symmetry analysis}
Considering the invariance of
equation (\ref{secondordercmee}) under a one parameter continuous Lie point symmetry group \cite{ibrahimov},
$x\rightarrow X=x+\epsilon\eta_1(t,x,y)+O(\epsilon^2)$,
$y\rightarrow Y=y+\epsilon\eta_2(t,x,y)+O(\epsilon^2)$,
$t\rightarrow T=x+\epsilon\xi(t,x,y)+O(\epsilon^2)$, $\epsilon<<1$, and
  performing the Lie point symmetry analysis
on the equation (\ref{secondordercmee}) using the package MULIE \cite{mulie}, we find that the system
admits only the following single Lie point symmetry vector,
\begin{eqnarray}
\Gamma_1=\frac{1}{2\lambda_1}\frac{\partial}{\partial t},
\end{eqnarray}
corresponding to the invariance of (\ref{secondordercmee}) under time translation for $\lambda_1\ne\lambda_2$.
For the specific choice $\lambda_1=\lambda_2$, which we have earlier proved to be superintegrable, we obtain
the following two additional point symmetries,
\begin{eqnarray}
&&\Gamma_2=-y\frac{k_1}{k_2}\frac{\partial}{\partial y}+y\frac{\partial}{\partial x},\nonumber\\
&&\Gamma_3=-x\frac{k_2}{k_1}\frac{\partial}{\partial x}+x\frac{\partial}{\partial y}.
\end{eqnarray}
So we conclude at this stage that since the system (\ref{secondordercmee}) is completely integrable and does not possess
enough number of Lie point symmetries, it has to
admit more general symmetries, namely nonlocal and contact symmetries.  The study requires separate analysis
and we do not pursue them here.  Note that on the other hand, the one dimensional MEE (\ref{1dmee}) admits
eight Lie point symmetries and is linearizable through point transformation which is not the case for the coupled
MEE (\ref{secondordercmee}).
\subsection{Painlev\'e singularity structure analysis}
We now perform the standard Painlev\'e singularity structure analysis \cite{ablowitz,sahadevan} to the equation (\ref{secondordercmee}).
Looking for the leading order behaviour of the Laurent series solution in the neighbourhood of a movable singular point $t_0$, we substitute
$x=a_0\tau^p$ and $y=b_0\tau^q$, $\tau=(t-t_0)\rightarrow 0$,
 and obtain
 \begin{eqnarray}
&& \hspace{-2cm}a_0p(p-1)\tau^{p-2}+a_0^3k_1^2\tau^{3p}+3a_0^2k_1p\tau^{2p-1}+2a_0b_0k_2p\tau^{p+q-1}
+a_0b_0k_2q\tau^{p+q-1}\nonumber\\
&&+2a_0^2b_0k_1k_2\tau^{2p+q}+a_0b_0^2k_2^2\tau^{p+2q}=0,\\
&& \hspace{-2cm}b_0q(q-1)\tau^{q-2}+b_0^3k_2^2\tau^{3q}+a_0b_0k_1p\tau^{p+q-1}+2a_0b_0k_1q\tau^{p+q-1}
+a_0^2b_0k_1^2\tau^{2p+q}\nonumber\\
&&+3b_0^2k_2q\tau^{2q-1}+2a_0b_0^2k_1k_2\tau^{p+2q}=0.
 \end{eqnarray}
 Comparing the exponents of $\tau$, we find $p=-1$ and $q=-1$.  Substituting this
and simplifying we get
\begin{eqnarray}
&&(2a_0-3a_0^2k_1+a_0^3k_1^2-3a_0b_0k_2+2a_0^2b_0k_1k_2+a_0b_0^2k_2^2)\tau^{-3}=0,\nonumber\\
&&(2b_0-3a_0b_0k_1+a_0^2b_0k_1^2-3b_0^2k_2+2a_0b_0^2k_1k_2+b_0^3k_2^2)\tau^{-3}=0\nonumber.
\end{eqnarray}
Solving the above system of equation we find two possibilities for the leading
order coefficients as $a_0=(1-b_0k_2)/k_1$ and $a_0=(2-b_0k_2)/k_1$,
while $b_0$ is arbitrary. For the choice
$a_0=(1-b_0k_2)/k_1$
we identify that the resonances (that is powers at which arbitrary constants can enter) occur at $r=-1$, $r=0$, $r=1$ and $r=1$.
By proceeding with the full Laurent series one can show that in addition to $t_0$ and $b_0$ being arbitrary (corresponding to $r=-1$ and $r=0$),
$a_1$ and $b_1$ are also arbitrary corresponding to $r=1,1$, while all higher order coefficients in the Laurent series can be determined in terms of
the earlier ones.  Similarly we find that the Laurent series corresponding to the second value of $a_0=(2-b_0k_2)/k_1$ also does not admit any
movable critical singular point.  We find that the equation (\ref{secondordercmee})
passes the Painlev\'e test as expected.
%\begin{center}
\subsection{Perturbed system : Numerical analysis}
In order to understand the dynamics of the system in the neighbourhood of the integrable parametric regime, we  perturb the system
leading to the form
\begin{eqnarray}
&&\ddot{x}+2(k_1x+k_2y)\dot{x}+(k_1\dot{x}+k_2\dot{y})x+
(k_1x+k_2y)^2x+\lambda_1 x+\rho_1 xy=0,\nonumber\\
&&\ddot{y}+2(k_1x+k_2y)\dot{y}+(k_1\dot{x}+k_2\dot{y})y+
(k_1x+k_2y)^2y+\lambda_2 y+\rho_2 xy=0,\label{gencmee}
\end{eqnarray}
where $\rho_1,\,\rho_2$ are the strength of the perturbation.  We now numerically
solve equation (\ref{gencmee}) for three different values of $\rho_1$ and $\rho_2$ using the fourth order
Runge-Kutta method.  Of these three parametric choices, the first one $(\rho_1=\rho_2=0)$ corresponds to the
completely integrable equation (\ref{secondordercmee}) and the other two are the perturbed
form of (\ref{secondordercmee}), namely,
$\rho_1=\rho_2=0.5$ and $\rho_1=\rho_2=0.7$.
  In figures 6a we show the phase space plots of equation (\ref{gencmee}) for these three different
parametric choices.  The corresponding Poincar\'e surface of sections are plotted in figures 6b by identifying the peaks of $y(t)$ from the time series data.
We find from the Poincar\'e surface of section of the torus that for the integrable parametric choice a closed curve
is obtained while as the strength of the perturbation is increased the curve starts to break up and ends up into scattered points, confirming the onset of chaos.

\begin{figure}[!ht]
\begin{center}
\includegraphics[width=.8\linewidth]{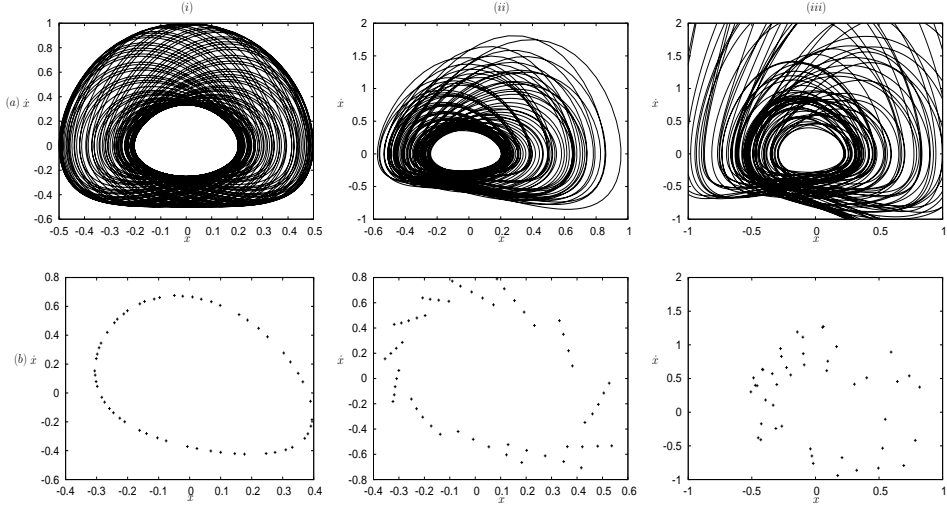}
\caption{Figures (a) (i), (ii) and (iii) describe phase space trajectories of equation (\ref{gencmee}) with $\lambda_1=1$ and $\lambda_2=2$ for three
values of $\rho_1,\,\rho_2$ corresponding to integrable case (i) $\rho_1=\rho_2=0$ and perturbed cases :
(ii) $\rho_1=\rho_2=0.5$ and (iii) $\rho_1=\rho_2=0.7$.
Figures (b) (i), (ii) and (iii) describe the Poincar\'e surface of sections of the above phase space plots.}
\end{center}
%\label{fig6}
\end{figure}

\section{$N$-coupled MEE}
Proceeding further, we find that equation (\ref{1dmee}) can also be generalized to arbitrary number, $N$, of coupled oscillators.
In order to generalize the results of two-coupled MEE (\ref{secondordercmee}) to $N$-coupled MEE, we have first investigated
 the three-coupled MEEs.  Then from results of two and three-coupled MEE, we have generalized the results to $N$-coupled MEE.
For brevity, we are not presenting here the results of the $N=3$ case.

The $N$-coupled modified Emden type equation is given as
\begin{eqnarray}
&&\hspace{-1.3cm}\ddot{x}_i+ 2(\sum_{j=1}^Nk_jx_j)\dot{x}_i+(\sum_{j=1}^Nk_j\dot{x}_j )x_i
+(\sum_{j=1}^Nk_jx_j)^2x_i+\lambda_ix_i=0,\,\,\,\,i=1,2,...,N.\label{ndim}
\end{eqnarray}
By generalizing the results of two and three-coupled cases one can now inductively construct the integrals of motion
for the equation (\ref{ndim}) which turn out to be
\begin{eqnarray}
&&\hspace{-1.5cm}I_{1i}=\frac{(\dot{x}_i+\sum_{j=1}^N(k_jx_j)x_i)^2+\lambda_ix_i^2}
{\left[\sum_{j=1}^N\left[\frac{k_j}{\lambda_j}(\dot{x}_j+\sum_{n=1}^N(k_nx_n)x_j)\right]+1\right]^2},\label{ndtidint}\\
&&\hspace{-1.5cm}I_{2i}=\tan^{-1}\left[\frac{\sqrt{\lambda_i}x_i}
{\dot{x}_i+\sum_{j=1}^N(k_jx_j)x_i}\right]-\sqrt{\lambda_i} t,\,\,\,\lambda_i>0,\quad i=1,2,...,N.\label{ndtdint}
\end{eqnarray}
In other words one has $N$ time-independent integrals and $N$ time-dependent integrals.  From these integrals one can
derive the general solution
 of equation (\ref{ndim}) for the parametric choice $\lambda_1,\lambda_2,...,\lambda_N>0$ as
\begin{eqnarray}
&&\hspace{-1cm}x_i(t)=\frac{A_i\sin(\omega_i t+\delta_i)}
{1-\sum_{j=1}^N\frac{A_jk_j}{\omega_j}\cos(\omega_jt+\delta_j)},\,\,\,i=1,...,N,\,\,
\bigg|\sum_{j=1}^N\frac{A_jk_j}{\omega_j}\bigg|<1,\label{ndimsol}
\end{eqnarray}
with $\omega_i=\sqrt{\lambda_i}$.

In the second case, $\lambda_i<0$, $i=1,2,...,N$, to derive the general solution we construct the time-dependent integrals in
the form
\begin{eqnarray}
&&\hspace{-1cm}I_{3i}=\frac{e^{2\sqrt{|\lambda_i|}t}(\dot{x}_i+\sum_{j=1}^N(k_jx_j)x_i-\sqrt{|\lambda_i|}x_i)}
{\dot{x}_i+\sum_{j=1}^N(k_jx_j)x_i+\sqrt{|\lambda_i|}x_i},\,\,i=1,2,...,N.
\end{eqnarray}
Using the above integrals one can construct the general solution of (\ref{ndim}) in the form
\begin{eqnarray}
&&\hspace{-1cm}x_i(t)=\frac{\prod_{s=1}^N\lambda_s\sqrt{|\lambda_i|}(\mathfrak{I}_{1i}
e^{\sqrt{|\lambda_i|}t}-\mathfrak{I}_{3i}e^{-\sqrt{|\lambda_i|}t})}
{\sum_{j=1}^N\prod_{m=1}^N\lambda_mk_j(\mathfrak{I}_{1j}e^{\sqrt{|\lambda_j|}t}+\mathfrak{I}_{3j}
e^{-\sqrt{|\lambda_j|}t})-2}, \,\,i=1,2,...,N,\label{gen_nd_sol}
\end{eqnarray}
where $j\ne m$ and $s\ne i$, $\mathfrak{I}_{1i}=\sqrt{I_{1i}}/(\sqrt{I_{3i}}\prod_{n=1}^N\lambda_n)$,\,
$\mathfrak{I}_{3i}=\sqrt{I_{1i}I_{3i}}/\prod_{n=1}^N\lambda_n$.

In the third case we consider the mixed sign case of the parameters $\lambda_i$, $i=1,2,...,N$.  Let us consider that
$\lambda_r$, $r=1,2,...,l$, are positive constants and $\lambda_j$, $j=N-l,...,N$, are negative constants.
The solution for this case can be written as
\begin{eqnarray}
x_r=-\omega_rA_r\sin(\omega_rt+\delta_r)h^{-1},\nonumber\\
x_j=\prod_{i=1}^l\omega_i^2\prod_{m=N-l}^N\lambda_m\sqrt{|\lambda_j|}
(\mathfrak{I}_{1j}e^{\sqrt{|\lambda_j|}t}
+\mathfrak{I}_{3j}e^{\sqrt{|\lambda_j|}t})(2h)^{-1},
\end{eqnarray}
where
\begin{eqnarray}
\hspace{-2.5cm}h=\sum_{r=1}^l k_rA_r\cos(\omega_r t+\delta_r)
+\sum_{m=N-l}^N k_m\prod_{r=1}^l\omega_r^2\prod_{q=N-l}^N\lambda_q(\mathfrak{I}_{1m}
e^{\sqrt{|\lambda_m|}t}+\mathfrak{I}_{3m}e^{-\sqrt{|\lambda_m|}t})-1,\nonumber
\end{eqnarray}
$m\ne q$ and $m\ne j$.

In the fourth case we have the $N$-dimensional generalization of ``classical MEE".  It is
evident from the lower dimensional cases that the integrals of motion includes (i) $(N-1)$ time-independent ones
(ii) $N$ time-dependent integrals which are linear in `$t$' and (iii) one time-dependent integral which is quadratic in
`$t$', that is
\begin{eqnarray}
&&I_{1i}=\frac{\dot{x}_i+x_i\sum_{j=1}^{N}k_jx_j}{\dot{x}_N+x_N\sum_{j=1}^{N}k_jx_j},\qquad i=1,2,\ldots, N-1,\\
&&I_{2i}=-t+\frac{x_i}{\dot{x}_i+x_i\sum_{j=1}^{N}k_jx_j},\;\; i=1,2,\ldots, N,\\
&&I_{3}=\frac{t^2}{2}+\frac{1-t\sum_{j=1}^{N}k_jx_j}{(\sum_{j=1}^{N}k_j\dot{x_j}+(k_jx_j)^2)}.
\end{eqnarray}
From these integrals, again one can deduce the general solution of (\ref{ndim}) for $\lambda_i=0$ in the form
\begin{eqnarray}
&&x_i(t)=\frac{2I_{1i}(I_{2i}+t)}
{\sum_{j=1}^{N}k_j I_{1j}(2I_{3}+(2I_{2j}+t)t)},\quad i=1,2,\ldots,N-1,\\
&&x_N(t)=\frac{2(I_{2N}+t)}
{\sum_{j=1}^{N}k_j I_{1j}(2I_{3}+(2I_{2j}+t)t)},
\end{eqnarray}
where $I_{1N}=1$.

In the fifth case, one can consider the parameters $\lambda_i$'s are all equal but nonzero, that is
$\lambda_1=\lambda_2=...=\lambda_N=\lambda\ne0$.
In this case, one can construct $(N-1)$ additional time-independent integrals by
eliminating the variable `$t$' in equation (\ref{ndtdint}),
\begin{eqnarray}
&&\hspace{-1.5cm}I_{3i}=\frac{(x_i\dot{x}_{i+1}-x_{i+1}\dot{x}_i)}
{\sum_j(k_j\dot{x}_j+(k_jx_j)^2+\lambda)},\,\,\, i=1,2,...N-1.\label{ndsuperint}
\end{eqnarray}
Again the existence of $(2N-1)$ time-independent integrals of motion (vide equations (\ref{ndtidint}) and (\ref{ndsuperint}))
confirms that the system
under consideration, namely (\ref{ndim}) with $\lambda_i=0$, $i=1,2,...,N$, is a maximally superintegrable one.
\section{Connection to uncoupled harmonic oscillators}
The solution of the coupled MEE equation (\ref{secondordercmee}) given by equation (\ref{cmee04b}) and (\ref{sol2})
can be rewritten as
\begin{eqnarray}
&&x=\frac{U}{1-\frac{k_1}{\omega_1^2}\dot{U}-\frac{k_2}{\omega_2^2}\dot{V}},\,\,
y=\frac{V}{1-\frac{k_1}{\omega_1^2}\dot{U}-\frac{k_2}{\omega_2^2}\dot{V}},\label{soluv}
\end{eqnarray}
where $U=A\sin(\omega_1t+\delta_1)$ and $V=B\sin(\omega_2t+\delta_2)$.  Here $U$ and
$V$ can also be interpreted as the solutions of the following uncoupled harmonic oscillator equations,
\begin{eqnarray}
\ddot{U}+\omega_1^2U=0,\qquad\ddot{V}+\omega_2^2V=0.\label{uncoupsho}
\end{eqnarray}

Equation (\ref{soluv}) gives a transformation connecting
equation (\ref{secondordercmee}) and the harmonic oscillator equations (\ref{uncoupsho}).  In order to invert this
transformation,
we need $\dot{U}$ and $\dot{V}$ for which
we differentiate equation (\ref{soluv}) once with respect to time and replace $\ddot{U},\ddot{V}$
with $-\omega_1^2U$ and $-\omega_2^2V$ respectively.  Thus we get the following equations
\begin{eqnarray}
&&\hspace{-2.1cm}\dot{x}=\frac{\dot{U}(1-\frac{k_1\dot{U}}{\omega_1^2}-\frac{k_2\dot{V}}
{\omega_2^2})-U(k_1U+k_2V)}
{\left(\frac{k_1\dot{U}}{\omega_1^2}+\frac{k_2\dot{V}}{\omega_2^2}-1\right)^2},\,\,\,
\dot{y}=\frac{\dot{V}(1-\frac{k_1\dot{U}}{\omega_1^2}-\frac{k_2\dot{V}}
{\omega_2^2})-V(k_1U+k_2V)}
{\left(\frac{k_1\dot{U}}{\omega_1^2}+\frac{k_2\dot{V}}{\omega_2^2}-1\right)^2}.\label{soluvdot}
\end{eqnarray}
Solving the relations (\ref{soluv}) and (\ref{soluvdot}) one obtains the inverse transformation of (\ref{soluv}) and (\ref{soluvdot}) as
\begin{eqnarray}
&&\hspace{-1.5cm}U=\frac{x}
{(\frac{k_1}{\omega_1^2}(\dot{x}+(k_1x+k_2y)x)+\frac{k_2}{\omega_2^2}(\dot{y}+(k_1x+k_2y)y)+1)},\label{u}\\
&&\hspace{-1.5cm}V=\frac{y}
{(\frac{k_1}{\omega_1^2}(\dot{x}+(k_1x+k_2y)x)+\frac{k_2}{\omega_2^2}(\dot{y}+(k_1x+k_2y)y)+1)},\\
%\end{eqnarray}
%\begin{eqnarray}
&&\hspace{-1.5cm} \dot{U}=\frac{(\dot{x}+k_1x^2+k_2xy)}
{(\frac{k_1}{\omega_1^2}(\dot{x}+(k_1x+k_2y)x)+\frac{k_2}{\omega_2^2}(\dot{y}+(k_1x+k_2y)y)+1)},\\
 &&\hspace{-1.5cm} \dot{V}=\frac{(\dot{y}+k_1xy+k_2y^2)}
{(\frac{k_1}{\omega_1^2}(\dot{x}+(k_1x+k_2y)x)+\frac{k_2}{\omega_2^2}(\dot{y}+(k_1x+k_2y)y)+1)}.\label{vdot}
\end{eqnarray}

This is indeed a contact transformation between the old and new variables, which can also be interpreted as a linearizing transformation
to equation (\ref{secondordercmee}).
The form of the Hamiltonian for the system of
two uncoupled harmonic oscillators (\ref{uncoupsho}) obviously is
\begin{eqnarray}
H=\frac{1}{2}\left[P_1^2+P_2^2+\lambda_1U^2+\lambda_2V^2\right],
\end{eqnarray}
where $P_1=\dot{U}$, $P_2=\dot{V}$, $\lambda_1=\omega_1^2$ and $\lambda_2=\omega_2^2$.
Substituting for $P_1,\,P_2,\,U,$ and $V$ from (\ref{u})-(\ref{vdot}) we get,
\begin{eqnarray}
\hspace{-1.5cm}H=\frac{(\dot{x}+k_1x^2+k_2xy)^2+(\dot{y}+k_1xy+k_2y^2)^2+\lambda_1x^2+\lambda_2y^2}
{2[\frac{k_1}{\lambda_1}(\dot{x}+(k_1x+k_2y)x)+\frac{k_2}{\lambda_2}(\dot{y}+(k_1x+k_2y)y)+1]^2}.\label{harham}
\end{eqnarray}
Here we note that $H=\frac{1}{2}(I_1+I_2)$, where $I_1$ and $I_2$ are the time independent integrals of motion of equation (\ref{secondordercmee})
(vide  (\ref{2dtidint1}) and (\ref{2dtidint2})).

Further, $U$ and $V$ satisfy the canonical equations
\begin{eqnarray}
&&\dot{U}=\frac{\partial H}{\partial P_1}=P_1,\quad \dot{P_1}=-\frac{\partial H}{\partial U}=-\lambda_1U,\label{harcan1} \\
&&\dot{V}=\frac{\partial H}{\partial P_2}=P_2,\quad\dot{P_2}=-\frac{\partial H}{\partial V}=-\lambda_2V\label{harcan2}.
\end{eqnarray}
The results obviously confirm the existence of a conservative Hamiltonian for equation (\ref{secondordercmee}).

The above results can also be extended to $N$ dimensions by using the contact transformation derivable from the solution
(\ref{ndimsol}),
\begin{eqnarray}
&&U_i=\frac{x_i}{\sum_{j=1}^N\left[\frac{k_j}{\lambda_j}(\dot{x}_j+\sum_{n=1}^N(k_nx_n)x_j)\right]+1},\nonumber\\
&&P_i=\dot{U}_i=\frac{\dot{x}_i+(\sum_{j=1}^Nk_jx_j )x_i}
{\sum_{j=1}^N\left[\frac{k_j}{\lambda_j}(\dot{x}_j+\sum_{n=1}^N(k_nx_n)x_j)\right]+1},
\end{eqnarray}
 $i=1,...,N$. Consequently the Hamiltonian for equation (\ref{ndim}) can be rewritten as a system
 of $N$ uncoupled harmonic oscillators specified by
\begin{eqnarray}
H=\frac{1}{2}\sum_{i=1}^N\left(P_i^2+\lambda_iU_i^2\right).
\end{eqnarray}
In terms of the original coordinates this becomes
\begin{eqnarray}
H=\frac{(\dot{x}_i+(\sum_{j=1}^Nk_jx_j )x_i)^2+\lambda_i x_i^2}
{2\left[\sum_{j=1}^N\left[\frac{k_j}{\lambda_j}(\dot{x}_j+\sum_{n=1}^N(k_nx_n)x_j)\right]+1\right]^2},\label{ndimH}
\end{eqnarray}
which can be associated with the integrals of motion (\ref{ndtidint}) as $H=\frac{1}{2}\sum_{i=1}^N I_{1i}$.
The canonical equation of motion are given as
\begin{eqnarray}
\dot{U}_i=\frac{\partial H}{\partial P_i},\,\,\,\dot{P}_i=-\frac{\partial H}{\partial U_i},\,\,\,\,i=1,2,...,N,
\label{ncanhar}
\end{eqnarray}
thus confirming the Hamiltonian nature of the system (\ref{ndim}).
However,  we have not yet succeeded to obtain an explicit Lagrangian form, and so canonically
conjugate momenta, to re-express (\ref{harham}) or (\ref{ndimH})  in terms of canonical coordinates. This is being pursued at present.
\section{Conclusion}
In this paper we have presented a system of completely integrable $N$-coupled Li\'enard type (modified Emden type)
nonlinear oscillators.  The system admits in general $N$ time-independent and $N$
time-dependent integrals whose explicit forms can also be found.  For special parametric choices, the
system also becomes maximally superintegrable.  Using these integrals general solution of
periodic, quasiperiodic, frontlike and decaying type or oscillatory type are obtained depending on
the signs and magnitudes of the linear forcing terms.  We have also pointed out that the
system possesses a nonstandard Hamiltonian structure and is transformable to a system of uncoupled
harmonic oscillators.  Further analysis of the Hamiltonian
structure can be expected to yield interesting information on the nonstandard Hamiltonian structure
of coupled nonlinear oscillators of dissipative type.  It is also of interest to investigate whether there
exist other couplings of MEEs  and its generalizations which are also integrable : For example, one can show
\cite{kundu} that under the general transformation $U(t,x)=x^ne^{\int_0^t f(x(t'))dt'}$, $f(x(t))=\beta(t)x^m+\gamma(t)$, the one dimensional generalized MEE,
$\ddot{x}+(n-1)\frac{\dot{x}^2}{x}+\frac{\beta^2}{n}x^{2m+1}+b_1(t,x)\dot{x}
+b_2(t)x^{m+1}+b_3(t)x=0 $, where $b_1(t,x)=\frac{1}{n}\left(2n\gamma+n\lambda+(m+2n)\beta x^m\right)$,
\,$b_2(t)=\frac{1}{n}\left(\dot{\beta}+2\gamma\beta+\lambda\beta\right)$,
$b_3(t)=\frac{1}{n}(\dot{\gamma}+\gamma^2+\lambda\gamma)$, can be reduced to the linear harmonic oscillator equation.
One can expect higher dimensional generalization of such a transformation can give rise to more general higher dimensional integrable equations.
  These questions are being pursued currently.
\ack
The work forms a part of a research project of MS and an IRHPA project of ML sponsored by the
Department of Science \& Technology (DST), Government of India.  ML is also supported by a DST Ramanna Fellowship.

\appendix
\section{Generalized modified Prelle-Singer procedure}
To solve the system of two-coupled second order nonlinear ODEs we apply the
generalized modified Prelle-Singer (PS) approach introduced recently \cite{royal1}.  Let
the system
\begin{eqnarray}
&&\ddot{x}=-2(k_1x+k_2y)\dot{x}-(k_1\dot{x}+k_2\dot{y})x-
(k_1x+k_2y)^2x-\lambda_1 x\equiv\phi_1,\nonumber\\
&&\ddot{y}=-2(k_1x+k_2y)\dot{y}-(k_1\dot{x}+k_2\dot{y})y-
(k_1x+k_2y)^2y-\lambda_2 y\equiv\phi_2,\nonumber
\end{eqnarray}
admits a
first integral of the form $I(t,x,y,\dot{x},\dot{y})=C$
 with $C$ constant on the solutions so that the total differential gives
\begin{eqnarray}
dI={I_t}{dt}+{I_{x}}{dx}+{I_{y}}{dy}+{I_{\dot{x}}}{d\dot{x}}
+{I_{\dot{y}}}{d\dot{y}}=0. \label {cso02}
\end{eqnarray}
Equation (\ref{secondordercmee}) can be rewritten as the equivalent 1-forms
\begin{eqnarray}
\phi_1dt-d\dot{x}=0,\qquad
\phi_2dt-d\dot{y}=0. \label {cso03}
\end{eqnarray}
Adding null terms
$s_1(t,x,y,\dot{x},\dot{y})\dot{x}dt
- s_1(t,x,y,\dot{x},\dot{y})dx$ and $s_2(t,x,y,\dot{x},\dot{y})\dot{y}dt
- s_2(t,x,y,\dot{x},\dot{y})dy$
 with the first equation in (\ref{cso03}), and
$u_1(t,x,y,\dot{x},\dot{y})\dot{x}dt  -
u_1(t,x,y,\dot{x},$ $\dot{y})dx $ and $u_2(t,x,y,\dot{x},\dot{y})\dot{y}dt -
u_2(t,x,y,\dot{x},\dot{y})dy $ with the second equation in (\ref{cso03}),
respectively, we obtain that, on the solutions, the 1-forms
%\numparts
\begin{eqnarray}
%\addtocounter{equation}{-1}
%\label {cso06}
%\addtocounter{equation}{1}
&&(\phi_1+s_1\dot{x}+s_2\dot{y})dt-s_1dx-s_2dy-d\dot{x}=0,\label {cso04}\\
&&(\phi_2+u_1\dot{x}+u_2\dot{y})dt-u_1dx-u_2dy-d\dot{y}=0.\label {cso05}
\end{eqnarray}
%\endnumparts

Hence, on the solutions, the 1-forms (\ref{cso02}) and
(\ref{cso04})-(\ref{cso05}) must be proportional. Multiplying (\ref{cso04}) by the function
$ R(t,x,y,\dot{x},\dot{y})$ and (\ref{cso05}) by
the function $ K(t,x,y,\dot{x},\dot{y})$, which act as the integrating
factors for (\ref{cso04}) and (\ref{cso05}), respectively, we have on the
solutions that
\begin{eqnarray}
\hspace{-1cm}dI=R(\phi_1+S\dot{x})dt+K(\phi_2+U\dot{y})dt-RSdx
-KUdy-Rd\dot{x}-Kd\dot{y}=0,\;\;\label {cso07}
\end{eqnarray}
where $S=(Rs_1+Ku_1)/R$ and $U=(Rs_2+Ku_2)/K$. Comparing
equations (\ref{cso07}) and (\ref{cso02}) we have, on the solutions,
the relations
\begin{eqnarray}
\hspace{-2cm} I_t  =R(\phi_1+S\dot{x})+K(\phi_2+U\dot{y}),\; I_{x}  = -RS,\,
  I_{y} =-KU,\;
 I_{\dot{x}}  =-R,\;
  I_{\dot{y}} =-K.
 \label {cso08}
\end{eqnarray}
The compatibility conditions between the different equations in (\ref{cso08}) provide us
the ten relations
\begin{eqnarray}
D[S] &=&-\phi_{1x}-\frac{K}{R} \phi_{2x}
    +\frac{K}{R}S\phi_{2\dot{x}}
           +S\phi_{1\dot{x}}+S^2,  \label {eq23}\\
D[U] &=&-\phi_{2y}-\frac{R}{K}\phi_{1y}
    +\frac{R}{K}U\phi_{1\dot{y}}
           +U\phi_{2\dot{y}}+U^2,  \label {eq24}\\
D[R]  &=&-(R\phi_{1\dot{x}}+K\phi_{2\dot{x}}+R S),  \label {eq25}\\
D[K]  &=&-(K\phi_{2\dot{y}}+R\phi_{1\dot{y}}+K U), \label {eq26}\\
SR_{y} &=&-RS_{y}+UK_{x}+KU_{x},\;\;R_{x} =SR_{\dot{x}}+RS_{\dot{x}},
  \label {eq31}\\
% \label {eq27}\\
R_{y} &=&UK_{\dot{x}}+KU_{\dot{x}},\qquad\qquad \;
%\label {eq28}\\
K_{x} =SR_{\dot{y}}+RS_{\dot{y}},  \label {eq29}\\
K_{y}&=&UK_{\dot{y}}+KU_{\dot{y}},\qquad\qquad\;
%\label {eq30}\\
R_{\dot{y}} =K_{\dot{x}}.\label {eq32}
\end{eqnarray}
Here the total differential operator, $D$, is defined by
$
D =\frac{\partial}{\partial{t}}+\dot{x}\frac{\partial}{\partial{x}}
+\dot{y}\frac{\partial}{\partial{y}}
+\phi_1\frac{\partial}{\partial{\dot{x}}}+\phi_2\frac{\partial}
{\partial{\dot{y}}}$.

Integrating equations (\ref{cso08}), we obtain the
integral of motion,
\begin{eqnarray}
I=r_1+r_2+r_3+r_4
-\int\bigg[K+\frac{d}{d\dot{y}}\bigg(r_1+r_2+r_3+r_4\bigg)\bigg]
d\dot{y},
\label {cso09}
\end{eqnarray}
where
\begin{eqnarray}
\hspace{-2cm}r_1&=\int\bigg(R(\phi_1+S\dot{x})+K(\phi_2+U\dot{y})\bigg)dt,
\quad
r_2=-\int\bigg(RS+\frac{d}{dx}(r_1)\bigg)dx,&\qquad\nonumber\\
\hspace{-2cm}r_3&=-\int\bigg(KU+\frac{d}{dy}(r_1+r_2)\bigg)dy,\qquad\quad\,\,
r_4=-\int\bigg[R+\frac{d}{d\dot{x}}\bigg(r_1+r_2+r_3\bigg)\bigg]d\dot{x}.&
\nonumber
\end{eqnarray}

Solving equations (\ref{eq23})-(\ref{eq32}) one can obtain $S,U,R$ and $K$.  Substituting these
forms into (\ref{cso09}) and evaluating the resulting integrals one can get the associated integrals of motion.  Once
sufficient number of integrals of motion are found (four in the present problem) then the
general solution can be derived from these integrals by just algebraic manipulations.  One can
refer to \cite{royal1} for details of the
method of solving the determining equations (\ref{eq23})-(\ref{eq32}).

\end{document}